\documentclass[tightenlines,nofootinbib,11pt,notitlepage]{revtex4-1}

\usepackage{graphicx}

\usepackage{subfigure}
\newcommand{\UFive}{{$^{235}$U}}
\newcommand{\UEight}{{$^{238}$U}}
\newcommand{\PuNine}{{$^{239}$Pu}}
\newcommand{\PuOne}{{$^{241}$Pu}}

\begin{document}

\title{Reactor antineutrino fluxes -- status and challenges}

\author{Patrick Huber}
\email{pahuber@vt.edu}
\affiliation{Center for Neutrino Physics, Virginia Tech, Blacksburg, VA}

\begin{abstract}
In this contribution we describe the current understanding of reactor
antineutrino fluxes and point out some recent developments. This is
not intended to be a complete review of this vast topic but merely a
selection of observations and remarks, which despite their
incompleteness, will highlight the status and the challenges of this
field.
\end{abstract}

\date{\today}

\maketitle

\section{Introduction}

The antineutrino flux from a nuclear reactor has become a matter of
considerable interest over the past few years. The antineutrinos are
created in the beta decay of the neutron rich isotopes produced as
fragments in the fission of the reactor fuel. The interest in the
resulting electron antineutrino flux originates from two
communities. Basic research employs measurement of the flux to
investigate neutrino oscillations including the possible existence of
sterile neutrinos while the safeguards and threat reduction community
would use the neutrino\footnote{For the sake of brevity we will refer
  to electron antineutrinos as neutrinos throughout this paper.}
spectrum and its composition over time as indicator of the makeup of the
fissile material in the reactor.  The basic research focus is on the
absolute neutrino flux while safeguards has greater interest in the
spectrum shape which may have markers for particular species of the
fuel. Significant uncertainty remains regarding both issues.
Reactor neutrino experiments rely on inverse beta decay (IBD)
\begin{equation}
\bar\nu_e + p\rightarrow n + e^+
\end{equation}
to detect the neutrino. This reaction has a neutrino energy threshold
of $E_\mathrm{th}simeq 1.8\,MeV$. Any uncertainty in the cross section
directly relates to an uncertainty in the detected event rate or
measured flux.  The existing world average of the absolute value of
the measured flux is ~6\% below the best prediction of that
flux~\cite{Mention:2011rk}, which was recently confirmed by Daya
Bay~\cite{An:2015nua}, which known as the Reactor Antineutrino Anomaly
(RAA).

Three recent very successful experiments (Daya Bay~\cite{An:2012eh},
Reno~\cite{Ahn:2012nd}, Double Chooz~\cite{Abe:2011fz}) focused on
measuring the neutrino mixing angle $\theta_{13}$ and as a by-product
provided the most precise and detailed measurements of the neutrino
spectrum produced by pressurized water power reactors (PWR). All three
measurements used well-calibrated detectors at three different reactor
sites and observed an unexpected excess of neutrinos with energies
between 4.8 and 7.3 MeV~\cite{An:2015nua}. This result has forcefully
brought home the notion that the neutrino fluxes are not as well
understood as had been thought. At present, it is not clear what
physics gives rise to the bump. It clearly must be attributed to the
excess production of some isotope or isotopes with a beta decay end
point energy in the interval of the observed bump.  There was a belief
that the reactor neutrino fluxes could be predicted to within
2\%. This belief was founded on employing integral beta spectra
measured in the 1980’s at the Institute Laue-Langevin (ILL) by
K. Schreckenbach and collaborators. They inserted foils of \UFive,
{\PuNine} and {\PuOne} into the ILL reactor to expose them to a
thermal neutron flux and directly measured an integral beta spectrum
created by the beta decaying isotopes produced by the neutron induced
fission of each fissile
isotope~\cite{VonFeilitzsch:1982jw,Schreckenbach:1985ep,Hahn:1989zr}. The
beta electron spectroscopy was performed with a magnetic spectrometer
which also provide the necessary electron/gamma separation.  It is of
note that this type of measurement has been pioneered by Reines in
1958~\cite{Carter:1959qm} using an anti-coincidence counter based on
plastic scintillator; the same technique was employed in a recent
measurement of the integral beta spectrum of
{\UEight}~\cite{Haag:2013raa}. Those measurements and the inferred
neutrino yield for each fissile isotopes can be combined with the
evolution of the fissile fuel composition over the run time to make a
prediction of the neutrino spectrum.  Of course, there are some
assumptions and physics required in going from the beta spectrum to
the neutrino spectrum but these were presumed to be tractable. Thus,
the bump observed in the neutrino flux came as a surprise as no such
bump could be generated using the ILL beta spectrum measurements. In
principle one could take a different tack from using the ILL
measurements and employ information contained in the very large data
bases ENDF/B-VII.1 and JEFF-3.1.1 associated with the fission of
{\UFive}, {\PuNine} and \PuOne. These databases pull together a large
body of experimental results to establish the fraction of each isotope
produced in the fission of a specific fuel element as well as the
subsequent beta decay branching ratios for each isotope. Naturally the
use of such a procedure produces a large uncertainty in the predicted
neutrino spectrum, on the order of 15\%, see for instance
Ref.~\cite{Fallot:2012jv}. However using ENDF/ B-VII.1 one predicts a
bump similar to that observed in the neutrino measurements. As
reported in~\cite{Hayes:2015yka} using the JEFF-3.1.1 no such bump is
predicted.  Reference~\cite{Hayes:2015yka} discusses possible origins
for the bump but provides no definite conclusions. Thus the bump in
the neutrino energy spectrum that at present cannot be traced to a
particular fissile isotope and an apparent 6\% deficit in the total
measured rate present serious obstacles to the use of neutrino
detection for either basic research or threat reduction.

For basic
neutrino research, the question is whether the 6\% deficit is due to nuclear
physics or due neutrino oscillation involving one or several eV-scale
sterile neutrinos. The eV-scale sterile neutrino interpretation is
also supported by a range of anomalies, where none taken individually
is statistically very significant, but which in combination point
towards an eV-scale sterile neutrino, for a review see
Ref.~\cite{Abazajian:2012ys}. 

The basic application for threat reduction and nuclear
non-proliferation safeguards relies on the predicted spectral
difference in neutrino emission between uranium and plutonium
isotopes, which allows to infer the plutonium content of a reactor
without reference to its past operating history and without the need
to modify reactor operations~\cite{Christensen:2014pva}. The basic
concept has been proposed by Borovoi and Mikaelyan in
1978~\cite{Borovoi:1978} and in many past reactor experiments a clear
correlation between the neutrino signal and the state of the reactor
was found (for early results
see~\cite{Korovkin:1988,Klimov:1990,Bernstein:2008tj}). For an actual
real-word application a much better quantitative understanding of the
spectral differences in neutrino yields between different fissile
isotopes is required.

The field of geoneutrino research as an experimental science is quite
young~\cite{Araki:2005qa} and has made significant progress in the
past few years~\cite{Gando:2013nba,Agostini:2015cba}. The sources and
distribution of heat in the Earth interior is an important question in
geophysics as it closely relates to the composition of the Earth and
it is this heat which drives plate tectonics and the geodynamo. There
are basically three potential sources for heat inside the Earth:
contraction or gravitational binding energy, chemical energy and
radioactivity. The overwhelming majority of radiogenic heat stems from
the decay chains of potassium-40, uranium-238 and thorium-232. For a
review of the relation between radiogenic heat and Earth composition
models see for instance Ref.~\cite{Sramek:2013}. The latter two decay
chains in uranium-238 and thorium-232 produce neutrinos above the
inverse beta decay threshold and thus are solely responsible for the
observed signals mentioned above. The next crucial data set will come
from JUNO, however a large background of reactor neutrinos will have
to be accurately subtracted~\cite{Han:2015roa}. This subtraction
requires a very good understanding of in particular the low-energy
part of the reactor neutrino spectrum between IBD threshold and about
3.5\,MeV.

%%%%%%%%%%%%%%%%%%%%%%%%%%%%%%%%%%%%%%%%%%%%%%%%%%%%%%%%%%%%%%%%%%%%%%%%
\section{Neutrino yields}

More than 99\% of the power in reactors, in a uranium fuel cycle, is
produced in the fission of four isotopes: $^{235}$U, $^{239}$Pu,
$^{238}$U, and $^{241}$Pu. A reactor with fresh fuel starts with only
fissions in the uranium isotopes and plutonium is produced via neutron
capture on $^{238}$U as the burn-up
increases. The total neutrino flux from a reactor $\phi$
can be written as
\begin{equation}
\phi(E)=\sum_I f_I S_I(E)\,,
\end{equation}
where $f_I$ is the fission rate in isotope $I$ and $S_I(E)$ is the
neutrino yield for the isotope $I$. The thermal power of the reactor
is also given in terms of the fission rates
\begin{equation}
\label{eq:power}
P_\mathrm{th}=\sum_I f_I p_I\,,
\end{equation}
where $p_I$ is the thermal energy release in one fission of the
isotope $I$; we use the values for $p_I$ given in
Ref.~\cite{Kopeikin:2004cn}.  In order to be able to disentangle the
contributions of the four isotopes, we need to know the neutrino yields
$S_I$. These neutrino yields, in principle, are given by the neutrino
spectra $\nu_k(E)$ of each fission fragment $k$ and the cumulative
fission yield for each fragment, $Y_k^I$,
\begin{equation}
S_I(E)=\sum_k Y_k^I \nu_k(E)\,,
\end{equation}
where $k$ typically runs over about 800 isotopes. In practice, we do
not know the neutrino spectrum of a given fission fragment, but have
only information regarding the beta spectrum and in many cases this
knowledge is inaccurate, incomplete, or entirely missing. Even for a
well known beta spectrum, significant complications arise from the
conversion of a beta spectrum into a neutrino spectrum since each
individual beta decay branch has to be treated separately. As a result,
a direct computation of the neutrino yields $S_I$ via the summation of
all individual neutrino spectra will be of limited
accuracy~\cite{Mueller:2011nm,Fallot:2012jv}, but in many cases is the
only available method.

A more accurate method is based on the measurement of the integral
beta spectrum of all fission
fragments~\cite{VonFeilitzsch:1982jw,Schreckenbach:1985ep,Hahn:1989zr,Haag:2013ra}
and subsequently the neutrino spectrum can be reconstructed from those
measurements~\cite{Huber:2011wv}. This method is less dependent on
nuclear data about individual fission fragments but is not entirely
free from uncertainties related to effects of nuclear
structure~\cite{Huber:2011wv,Hayes:2013wra}. In particular Hayes {\it
  et al.}~\cite{Hayes:2013wra} pointed out that forbidden decays which
can make up as much as 40\% of all neutrinos in certain energy
ranges can have a significant impact on the predictions. The reason is,
that in forbidden decays the spectrum of emitted neutrinos depends
on details of the underlying nuclear structure, and generally no
information at this level of detail is available.

Until the 2011 work by a group from Saclay~\cite{Mueller:2011nm}, the
results from
Refs.~\cite{VonFeilitzsch:1982jw,Schreckenbach:1985ep,Hahn:1989zr}
obtained in the 1980s at the Institut Laue-Langevin in Grenoble were
considered the gold standard. The Saclay group, in preparation of the
Double Chooz neutrino experiment~\cite{Abe:2011fz}, revisited the
previous results in an attempt to reduce the uncertainties. Instead,
they found a upward shift of the central value of the average yield by
about 3\% while the error budget remained largely unchanged. This
result, in turn, requires a reinterpretation of a large number of
previous reactor neutrino experiments, since this changes the expected
number of events. Together with the changes of the value of the
neutron lifetime~\cite{Greene} and corrections from so-called
non-equilibrium effects, the previous experiments appear to observe a
deficit in neutrino count rate of about 6\%; this is called the
reactor antineutrino anomaly and was first discussed in
Ref.~\cite{Mention:2011rk}. The initial result on the flux evaluation
and the 3\% upward shift has been independently
confirmed~\cite{Huber:2011wv}. A plausible explanation could come in
the form of a new particle, a sterile neutrino, which is not predicted
by the Standard Model of particle physics. Given the far-flung
consequences of the existence of this sterile neutrino a considerable
level of research activity ensued.

In Tab.~\ref{tab:fluxes} (taken from Ref.~\cite{Christensen:2013eza})
the event rate predictions for various flux models are compared for
the four fissile isotopes. The ENSDF flux model represents a crude
summation calculation and is based on thermal neutron fission yields
of $^{235}$U, $^{239}$Pu, and $^{241}$Pu from the JEFF database,
version~3.1.1~\cite{JEFF}; the fast neutron fission yield of $^{238}$U
from the ENDF-349 compilation~\cite{lanl}; and on the beta-decay
information contained in the Evaluated Nuclear Structure Data File
(ENSDF) database, version~VI~\cite{ensdf}. The neutrino spectrum is
derived following the prescription in Ref.~\cite{Huber:2011wv}. This
calculation reproduces the measured total beta
spectra~\cite{VonFeilitzsch:1982jw,Schreckenbach:1985ep,Hahn:1989zr,Haag:2013ra}
to within about 25\%. A detailed summation result has been derived by
Fallot {\it et al.}~\cite{Fallot:2012jv}, where the ENSDF entries are
replace with high quality experimental data (where available) and a a
selected mix of databases is used. Fallot's calculations reproduces
the measured total beta
spectra~\cite{VonFeilitzsch:1982jw,Schreckenbach:1985ep,Hahn:1989zr,Haag:2013ra}
to within 10\%. A direct inversion of the neutrino spectra from the
total beta data was performed in Ref.~\cite{Huber:2011wv} for the
isotopes $^{235}$U, $^{239}$Pu, and $^{241}$Pu. To date this
represents the most accurate neutrino yields for those isotopes. The
absolute values are significantly different between models, but once
normalized to the predictions for total rate and mean energy of
$^{235}$U, these results become very similar. Therefore, we conclude
that the difference in neutrino yield and mean energy between the
fissile isotopes is consistently predicted by the various flux models
-- which should come as no surprise since these differences have their
origin in the fission yields.
\begin{table}[t]
\begin{center}
\begin{tabular}{l|c|c|c|c|c|c|c|c|c|c|c|c|}
&\multicolumn{4}{c|}{ENSDF} & \multicolumn{4}{c|}{Fallot} & \multicolumn{4}{c}{Huber} \\ \cline{2-13}
& events & ratio & $\langle E\rangle [\mathrm{MeV}]$ & ratio & events & ratio & $\langle E\rangle [\mathrm{MeV}]$ & ratio & events & ratio &$\langle E\rangle [\mathrm{MeV}]$ & ratio  \\ \hline
 $^{235}$U & 3826 & 1 &4.48&1 & 3905 & 1 &4.28& 1& 4252 & 1 & 4.25 & 1 \\
 $^{238}$U & 5836 & 1.53 &4.59 & 1.024 & 6076 & 1.56 &4.45 & 1.040 &  &  & &  \\
 $^{239}$Pu & 2442 & 0.64 &4.26&0.950 & 2536 & 0.65 &4.13 &0.965 & 2796 & 0.66 & 4.04 & 0.951\\
 $^{241}$Pu & 3551 & 0.93&4.47&0.998 & 3515 & 0.90 &4.23&0.988& 3872 & 0.91 &4.13 & 0.971
\end{tabular}
\caption{\label{tab:fluxes} Rates and mean energies $\langle E \rangle
  $ for a 1\,MW$_\mathrm{th}$ reactor in a 1\,t detector at a standoff
  of 10\,m measuring for 1 year for each individual isotope, assuming
  that only this isotope is fissioning. The three different flux
  models are explained in the text. Ratios are given relative to
  $^{235}$U. From Ref.~\cite{Christensen:2013eza}.}
\end{center}
\end{table}

These results indicate a certain level of robustness in predictions,
but this impression needs to be tempered by the recent observation of
a bump-like feature. The shoulder recently observed in the neutrino
flux from PWRs as measured by Reno~\cite{Ahn:2012nd}, Daya
Bay~\cite{An:2012eh} and Double Chooz~\cite{Abe:2011fz} was unexpected
and its origin still uncertain. This shoulder cannot be reproduced if
one uses as input the Schreckenbach
measurements~\cite{VonFeilitzsch:1982jw,Schreckenbach:1985ep,Hahn:1989zr}
of the integral beta spectra of the daughters produced by thermal
neutron fission of reactor fuel. A subsequent publication by Dwyer and
Langford~\cite{Dwyer:2014eka} indicated that a shoulder similar to the
one observed could be produced using for input the beta decays in a
subset of ENDF/B-VII.1~\cite{ENDF} fission database. If one uses the
fission database JEFF-3.1.1~\cite{JEFF} as input no shoulder is
produced. This is not surprising as the uncertainties in the databases
are large relative to the size of the shoulder.

It would be useful if the observed shoulder could be uniquely assigned
to the decay of the daughters of a specific fuel type. If the shoulder
is due to the thermal neutron fission of \UFive, {\PuNine} or {\PuOne}
then Schreckenbach's measurements would be called into serious
question. The recent measurement of the beta spectrum of the decay of
the daughters of the fast fission of \UEight~\cite{Haag:2013ra} is not
sufficiently precise to show the presence of a shoulder.  Hayes and
collaborators~\cite{Hayes:2015yka} investigate the possible origins of
the observed shoulder and propose the following list
\begin{enumerate}
\item Beta decay of non-fissionable material in the reactor
\item Shape of the beta and neutrino spectrum for $\Delta J^{\Delta \pi}=0^-$ first forbidden decays
\item Beta decay of the daughters of the fast fission of \UEight  
\item Beta decay of daughters of the epithermal fission of \UFive, {\PuNine} and/or \PuOne
\item Errors in Schreckenbach's ILL beta spectra
\end{enumerate}

Taking at face value RENO's claim that the shoulder they observe makes
up 2\% of the total yield of events, allows the first of the proposed
causes to be readily dismissed. One shortcoming is, that the
modelers~\cite{Mueller:2011nm,Huber:2011wv} created the neutrino
spectrum from Schreckenbach’s beta spectrum assuming that all the beta
decays were allowed rather than taking account of the fact that some
of the most important decays are $\Delta J^{\Delta\pi}=0^-$, so-called
non-unique first forbidden decays. These decays have no weak magnetism
correction which increase their contribution relative to what the
modelers provided. This is because weak magnetism typically decreases
the antineutrino component of an allowed axial transition above half
of the end point energy. Treating these decays more correctly
increases their yield in the region of the bump~\cite{Hayes:2015yka}
by somewhat less than 1\% of the total yield so it cannot account for all
of the shoulder.  Not enough is known of the decay of the daughters
produced by the fast fission of {\UEight} so it certainly could
contribute to the shoulder. RENO observes the largest shoulder in the
neutrino flux and cites the largest contribution from the fission of
\UEight. To account for the entire shoulder the isotopes dominating
the shoulder region would have to be 4 times larger in JEFF-3.1.1 and
2 times larger in ENDF/B-VII.1~\cite{Hayes:2015yka}. Thus it appears
that 3) likely makes some contribution to the shoulder.

Schreckenbach's measurements were carried out using the thermal flux
of the ILL reactor while the 3 measurements observing the shoulder
were carried out at PWRs. Is it possible that the harder neutron flux
spectrum in a PWR relative to the one at ILL could produce more
fission products that create the shoulder? While there appear to be
large fluctuations~\cite{Cowan1,Cowan2,Cowan3,Cowan4} in the ratio of
symmetric to asymmetric fission the average over the epithermal
resonances is quite compatible with what is measured with a thermal
flux. The lone exception might be {\PuNine} that has an isolated and
prominent fission resonance at 0.3\,eV. Fission of this resonance must
play a larger role in the neutron spectrum of a power reactor than is
the case for fission at thermal energies. Thus 4) could make a
contribution to the bump.  The possibility of an error in the ILL beta
spectrum measurements must also be entertained. In the discussion of
possibility 2) it was pointed out that using the ILL beta spectrum
measurements and properly accounting for $\Delta J^{\Delta\pi}=0^-$
transitions can only account for half of the shoulder. This raises the
possibility that these measured beta spectra are not
correct. Certainly the measurements were not easy and the spectrometer
employed~\cite{Mampe} was complex. Further the signal to background in
bump region was ~2.5/1 and it is not clear how the background
subtraction was carried out. One should not dismiss the possibility of
error in the ILL beta spectra. A high statistics measurement of the
neutrino flux at a research reactor fueled with highly-enriched
uranium (HEU) will produce neutrinos only via the fission of {\UFive}
and should settle some of the issues raised above.

Assuming that the Daya Bay result on the bump holds, we can ask the
question, which fissile isotope does it come from? To demonstrate how
a multi-reactor deployment of a 5\,ton detector can elucidate this
question, we compare the following four types of reactors: a Daya Bay
like pressurized water reactor (DYB), a pressurized water reactor with
1/3 of reactor-grade MOX fuel (MOX3), a research reactor like the BR2
in Belgium running on highly enriched uranium (BR2) and a fast breeder
reactor like the Fast Breeder Test Reactor (FBTR) in India. For DYB
the fission fraction an reactor characteristics correspond to values
of the actual data taking period at Daya Bay~\cite{DYB14}, while for
the others, we used semi-realistic models in terms of reactor power,
reactor up-time and detector distance. Specifically, the fission
fractions in the four fissile isotopes for the 1/3 MOX are based on a
3D, pin-level 1/8-core simulation~\cite{Anna}. For BR2 we make the
simplifying assumption that all fissions take place in {\UFive} and we
have tested that a few percent of fission in other isotopes does not
change the results. For the FBTR we take the core simulation performed
for a full size Indian Fast Breeder from Ref.~\cite{Glasser:2007} as
proxy. We assume that there is no breeding blanket and we neglect
fission in the even plutonium isotopes which overall contribute about
5\% of fissions.  The reactor parameters and fission rates are
summarized in Tab.~\ref{tab:reactors}.

\begin{table}[b]
\begin{tabular}{c|rrr|llll|r}
Reactor&power&stand-off&duty factor& \UFive&\UEight&\PuNine&\PuOne&Events\\
&[MW]&[m]&& & & & &\\
\hline
DYB&2\,800&25&1&0.586&0.076&0.288&0.05&2,188,000\\
MOX3&3\,200&25&1&0.51&0.066& 0.39&0.031&2,402,000\\
BR2&60&5.5&0.4&1&0&0&0&297,000\\
FBTR&60&10&0.4&0.0093&0.10&0.71&0.11&95,000\\
\end{tabular}
\caption{\label{tab:reactors} Properties and fission fractions for a
  set of representative reactors. Event numbers given are based on a
  one year exposure of a 40\% efficient, 5 ton detector.}
\end{table}

\begin{figure}
\begin{minipage}[c]{0.53\textwidth}
\includegraphics[width=\textwidth]{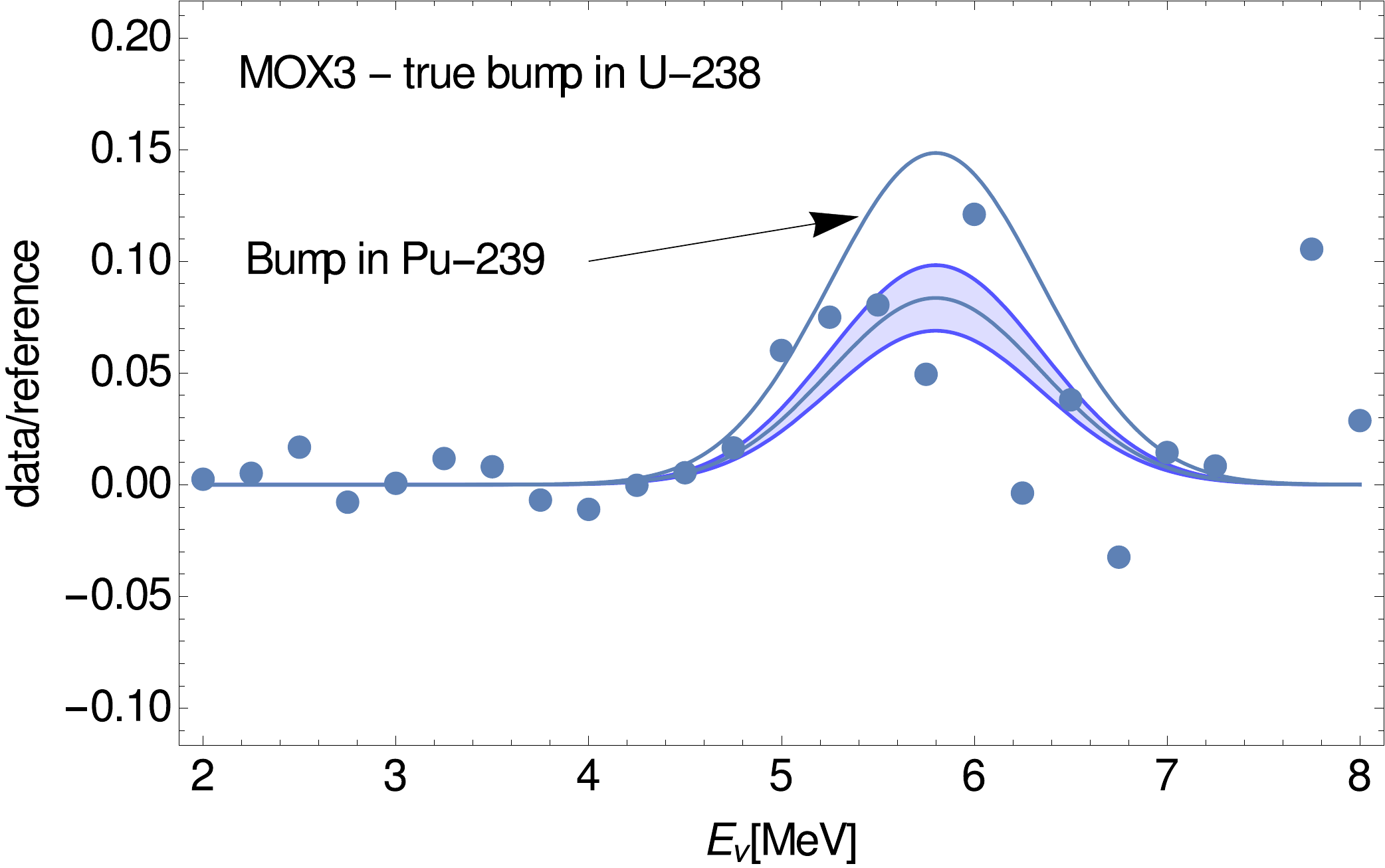}
\end{minipage}\hfill
\begin{minipage}[c]{0.45\textwidth}
\caption{\label{fig:mox3} Shown is the result for the ratio of data to
  prediction of a simulated experiment at MOX3 with the true bump in
  \UEight. This is based on a one year run of a 40\% efficient, 5 ton
  detector. The scatter in the simulated data (blue points) arises
  from a combination of statistical fluctuations and the systematic
  uncertainty of the underlying reference spectra.  The shaded region
  indicates the $1\,\sigma$ range from the fit and for comparison we
  show the expectation for MOX3 if the bump were in \PuNine.}
\end{minipage}
\end{figure}
For the following analysis we extract the shape, position and
amplitude of the bump from the Daya Bay data~\cite{DYB14}. We use as
reference spectrum for \UFive, {\PuNine} and {\PuOne} the ones from
Ref.~\cite{Huber:2011wv} and for {\UEight} the spectrum from
Ref.~\cite{Mueller:2011nm}. The bump appears relative to these
reference spectra and therefore any uncertainties in the reference
spectra itself will make it harder to detect the bump. We will use the
uncertainties as quoted in Ref.~\cite{Huber:2011wv} for \UFive,
{\PuNine} and {\PuOne} and assume a flat 10\% error for \UEight. Note,
that the reference spectra provide, except for the bump, an excellent
description of the Daya Bay data. We now can artificially choose to put
the bump into one isotope while ensuring to reproduce the right
amplitude in Daya Bay. We simulate data for a 5\,ton detector
with a 40\% detection efficiency and 1 year data taking. We 
impose random fluctuations for counting statistics and separately for
the underlying systematic uncertainty of the reference spectra. In
Fig.~\ref{fig:mox3} we show the result of one of these simulated
experiments at MOX3 with the true bump in \UEight. The shaded region
indicates the $1\,\sigma$ range from the fit and for comparison we
show the expectation for MOX3 if the bump were in \PuNine.

\begin{figure}[t]
\begin{minipage}[c]{0.53\textwidth}
\includegraphics[width=\textwidth]{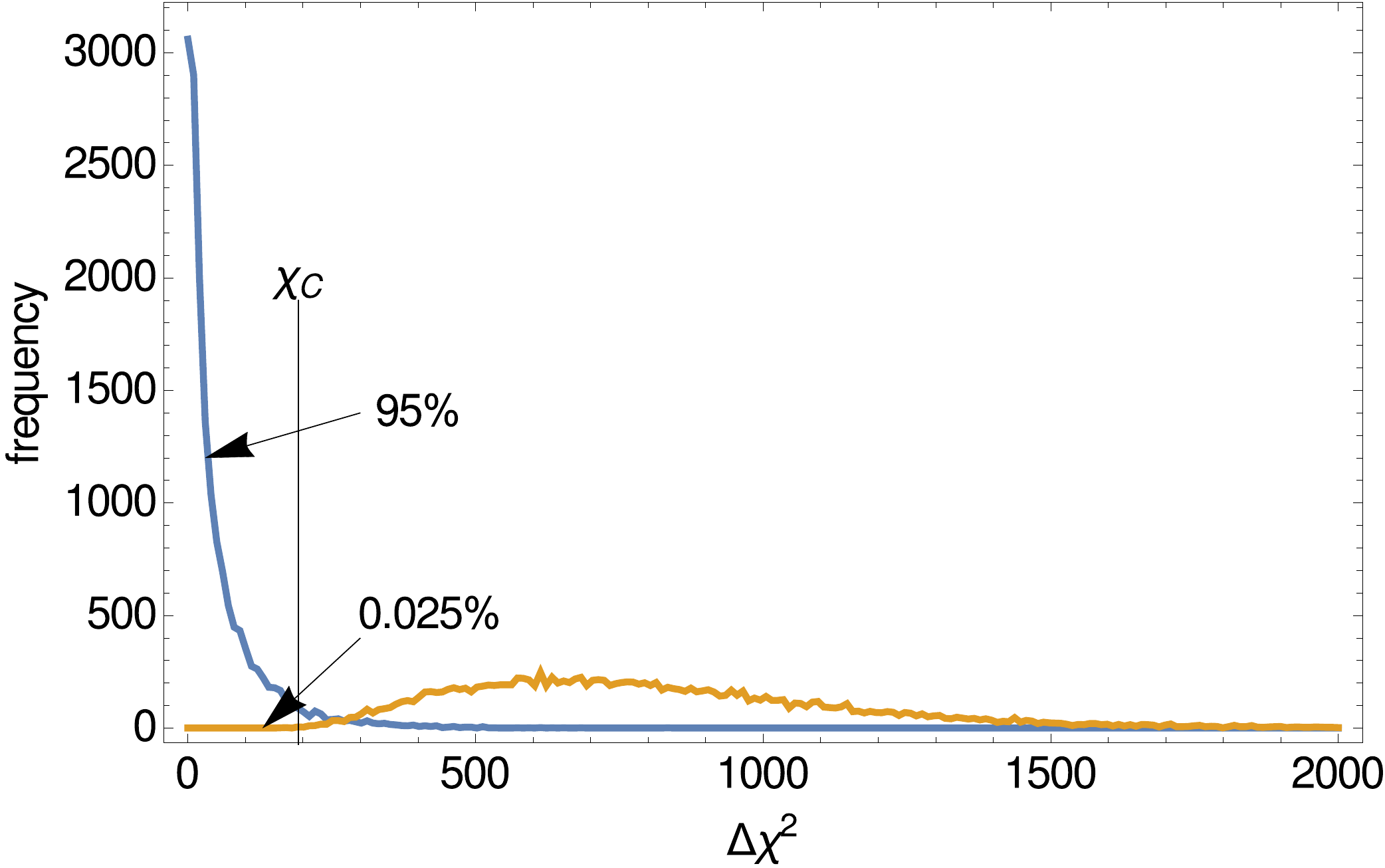}
\end{minipage}\hfill
\begin{minipage}[c]{0.45\textwidth}
\caption{\label{fig:histo}Shown is the $\Delta\chi^2$ distribution for
  the true bump in \UEight. The blue histogram shows the result if we
  fit the data assuming that the bump is in \UEight.  The orange
  histogram is obtained by fitting this data with the bump being in
  \PuNine. $\chi_c$ is defined by requiring that 95\% of all cases in
  the blue histogram are below this value. These results are based on
  a one year exposure of a 40\% efficient, 5 ton detector.}
\end{minipage}
\end{figure}
 We then in turn fit this data with the bump being in {\UFive} and
 then in {\UEight} {\it etc.} and compute a $\chi^2$ difference, where
 we leave the total event rate as a free parameter. This exercise is
 repeated 16,000 times, and we thus obtain a distributions for the
 $\chi^2$ and their differences $\Delta\chi^2$. For instance in
 Fig.~\ref{fig:histo} we show the $\Delta\chi^2$ distribution for the
 true bump in {\UEight} and the blue histogram is the result if we fit
 the data with the correct bump. The reason for the non-zero
 $\Delta\chi^2$ in this case is the systematic uncertainty of the
 reference spectra. The orange histogram is obtained by fitting this
 data with the bump being in \PuNine. $\chi_c$ is defined by requiring
 that 95\% of all cases in the blue histogram are below this
 value\footnote{In other words, for this hypothesis test, we set the
   error of the 1st kind to 5\%, that is we reject the true null
   hypothesis in 5\% of all cases.}, in this example $\chi_c=191.8$.
 Next, we look how many cases in the orange histogram also fall below
 $\chi_c$\footnote{This corresponds to the error of the 2nd kind, that
   is we accept the null hypothesis although it is wrong. The
   complement of this number corresponds to the power of the test.},
 which in this example are 4 out of 16,000 or 0.025\%. That is in only
 0.025\% cases we would conclude (wrongly) that {\PuNine} contains the
 bump, whereas in 95\% of the cases we would conclude (correctly) that
 the bump is in \UEight. In other words with an efficiency of 95\% we
 can reject the bump being in {\PuNine} at 3.67 standard deviations.
\begin{table}[b]
\begin{tabular}{|c|cccc|}
\hline
\multicolumn{5}{|c|}{MOX3}\\
\hline
Fit/True&\UFive&\UEight&\PuNine&\PuOne\\
\hline
\UFive& - & $> 4$  & $>4$  & $>4$ \\
\UEight& $>4$  & - & 3.8 & 0.6 \\
\PuNine& $>4$  & 3.7& - & $>4$  \\
\PuOne& $>4$  & 0.7 & $>4$  & - \\
\hline
\end{tabular}~\begin{tabular}{|c|cccc|}
\hline
\multicolumn{5}{|c|}{FBTR}\\
\hline
Fit/True&\UFive&\UEight&\PuNine&\PuOne\\
\hline
\UFive& - & $> 4$  & $>4$  & $>4$ \\
\UEight& $>4$  & - & 3.8 & 1.1\\
\PuNine& $>4$  & 3.6& - & $>4$  \\
\PuOne& $>4$  & 1.1 & $>4$  & - \\
\hline
\end{tabular}
\caption{\label{tab:sigs} Number of standard deviations at which for a
  given true bump a given fitted bump can be rejected while maintained
  a 95\% acceptance. Note, the MOX3 and FBTR data is correlated due
  the underlying common reference fluxes and hence standard deviations
  can not be added in quadrature.  These results are based on
  a one year exposure of a 40\% efficient, 5 ton detector.}
\end{table}

We repeat this exercise for each type of reactor and all 16
combinations of true and fitted bump being in a given
isotope. Clearly, DYB will see the same signal no matter which isotope
contains the bump as per definition of this analysis, it serves to
set the size and position of the bump.  The BR2 will see a very
strong signal for a bump ($>4\,\sigma$) if the bump is in {\UFive} and
see no bump if it is in any of the other isotopes. To diagnose the
case where the bump is  in {\UEight} or either in {\PuNine} or in {\PuOne} requires
reactors with increased plutonium and/or {\UEight} fission fractions
like MOX3 and FBTR. The resulting rejection power is shown in
Tab.~\ref{tab:sigs}.

All cases can be identified with better than 3 standard deviations
except the distinction between the bump being in {\UEight} versus
\PuOne, where the combination MOX3 and FBTR can reach about
$1.3\,\sigma$. Note that at this point this analysis is limited by the
systematic uncertainty of the underlying reference spectra. However,
the very same measurements will be able to deliver very precise new
reference spectra and thus the ultimate sensitivity will be
significantly higher. However, this analysis requires a very detailed
reactor modeling to determine the uncertainties of the fission
fractions. There are examples in the literature, {\it e.g.}
Ref.~\cite{Jones:2011hi}, and a similar exercise needs to be repeated
for the specific reactor in question. It also obvious that at least one 
 reactor with a high plutonium content needs to be added.

A similar analysis was presented recently in Ref.~\cite{Buck:2015clx}
where the focus was on a comparison of regular PWRs and research
reactors running on HEU, providing a clean {\UFive} signal. The results
presented here agree reasonably well with those of
Ref.~\cite{Buck:2015clx}, but we also show that reactor with very
different plutonium concentrations will be required to untangle the
bump. A major technical difference between the two analyses is that
here we fully account for the \emph{current} uncertainties of the
reference spectra, whereas in Ref.~\cite{Buck:2015clx} a smooth
interpolation through the bump region is used, which is equivalent to
assuming that reference spectra will have very significantly improved
by the time this measurement is perform. Thus, in reality the
sensitivity will be in-between the results of those two analyses.

%%%%%%%%%%%%%%%%%%%%%%%%%%%%%%%%%%%%%%%%%%%%%%%%%%%%%%%%%%%%%%%%
\section{Non-linear effects in reactor fluxes}

In Ref.~\cite{Huber:2015ouo} the effect of neutron capture isotopes on
the antineutrino spectrum is investigated and corrections of up to 1\%
for PWRs and several per cent for naval reactors are found in the
low-energy neutrino flux. Candidate isotopes can be found by
looking at isotopes which can undergo (two neutrino) double-beta decay:
fission fragments are generally produced a few beta decays away from
stability and they will decay in their mass chain down to the first
stable isotope they encounter, e.g. for the $A=100$ mass chain this
will be $^{100}$Mo. The next isotope in this chain is $^{100}$Tc which
itself beta decays with an endpoint of 3.2\,MeV, well above IBD
threshold, however it can not be produced by beta decay of $^{100}$Mo
due to nucleon-pairing effects, that is why $^{100}$Mo only double-beta
decays. Thus production of $^{100}$Tc via beta decay is impossible and
its direct fission yield is negligible. However, $^{99}$Tc is not
blocked by a double-beta decay isotope and thus is produced as a
result of beta decays along the $A=99$ mass chain. Again thanks to
pairing effects, $^{99}$Tc has a sizable neutron capture cross section
of about 17\,b which yields $^{100}$Tc, which in turn contributes to
the low-energy end of the neutrino spectrum. Under some simplifying
assumption the rate of $^{99}$Tc production is proportional to the
neutron flux, $\Phi_n$, but the capture rate to $^{100}$Tc is
proportional to neutron flux and the rate of $^{99}$Tc production and
as a result the rate of $^{100}$Tc production is proportional to
$\Phi_n^2$, that is it has a \emph{non-linear} dependence on the
neutron flux in contrast to regular fission fragments which have a
linear dependence. There is a simple analytic theory for the size of
the resulting correction, however this is accurate only within 
50\%. For a more precise calculation a detailed reactor burn-up
calculation is required and these detailed results are presented in
Ref.~\cite{Huber:2015ouo}.

%%%%%%%%%%%%%%%%%%%%%%%%%%%%%%%%%%%%%%%%%%%%%%%%%%%%%%%%%%%%%%%%%%%
\section{Summary}

Nuclear reactors have been the workhorse of neutrino physics from it's
very beginning as an experimental science~\cite{Cowan:1992xc} and much
has been learned about neutrino properties from a series of
experiments spanning many decades. Recently, a very precise
determination of $\theta_{13}$ has been achieved by using reactors as
a neutrino source and employing the comparison of data obtained with
near and far detectors, which essentially obviates the need to
understand the reactor neutrino flux.

Till 2011 reactor antineutrino fluxes appeared to be well understood
at the level of about 2\% uncertainty, but as outlined here and
elsewhere, this confidence was mistaken. As often with complex
problems, the closer one looks the larger the uncertainty
becomes. Predicting the inverse beta decay event rate with a reactor
as neutrino source is extraordinarily complex as it requires a
quantitative understanding of reactor physics to determine the neutron
flux and fission rates. From this information together with the
fission yields the isotopic composition of the reactor needs to be
determined. For each isotope a detailed understanding of its various
beta-decay branches is required and since about 30--40\% of all
neutrinos in the relevant energy regime are from forbidden decays, the
details of nuclear structure can not be avoided.   Also, there is a
number of low-energy effects related to isotopes which have
comparatively long half-lives giving rise the non-equilibrium
correction. These same isotopes also contribute to neutrino emissions
from spent nuclear fuel, which, if spent fuel is present on site, have
to be accounted for. More recently also non-linear effects in form of
neutron capture isotopes have been pointed out which will greatly
complicate the comparison of data from different reactors.

The precise measurements obtained at the near detectors of several
experiments also clearly highlight the limitations of our
understanding of reactor neutrino fluxes: the 5\,MeV bump remains a
conundrum. We explored certain experimental tests which could be
performed as was done in Ref.~\cite{Buck:2015clx} and it is clear that
even just assigning the responsible fissile isotope requires a
continued effort. The prediction of non-linear isotopes can be
verified by measuring the abundance of the stable end-point
isotopes. A series of close-range reactor measurements is planned,
which will add further information about reactor antineutrino fluxes,
but is worthwhile to point out that with the Daya Bay data set a very
precise measurement is available. The vicissitudes encountered close
to a reactor, that is a lack of overburden and reactor related
backgrounds, will make it a challenge to achieve comparable precision.

A central role in this tale is played by the beta spectrum
measurements performed by Schreckenbach {\it et al.} in the
1980s. They constitute the single point of failure for many
predictions and thus the question is: Can these measurements be
reproduced with similar precision? We did not touch on efforts to
improve the data on beta decays of the individual isotopes by using
totally active gamma spectroscopy, see for instance
Ref.~\cite{Zakari-Issoufou:2015vvp}, or the uncertainties inherent in
fission yields. Efforts to improve nuclear data bases will be very
beneficial to the issues outlined here. It will require a broad and
sustained effort by many communities to unravel the riddle of the
reactor neutrino flux, with potentially large discoveries to be made.

%%%%%%%%%%%%%%%%%%%%%%%%%%%%%%%%%%%%%%%%%%%%%%%%%%%%%%%%%%%%%%%%%%%%
\acknowledgements 

I would like to thank G.~Garvey for collaboration on an early version
of this manuscript and A. Erickson for providing the fission rates for
the MOX case studied here.  This work was in part supported by the
U.S. Department of Energy under award \protect{DE-SC0013632}.

%%%%%%%%%%%%%%%%%%%%%%%%%%%%%%%%%%%%%%%%%%%%%%%%%%%%%%%%%%%%%%%%%%%
%\bibliographystyle{apsrev} \bibliography{references}

%%%%%%%%%%%%%%%%%%%%%%%%%%%%%%%%%%%%%%%%%%%%%%%%%%%%%%%%%%%%%%%%%%%
\end{document}